\documentclass[superscriptaddress,twocolumn,amsmath,amssymb,footinbib]{revtex4-2}
\usepackage[english]{babel}
\usepackage[pdftex]{graphicx}
\usepackage{gensymb}
\usepackage[symbol]{footmisc}
\usepackage{xcolor}
\usepackage{csquotes}
\usepackage{mathrsfs}
\usepackage{subcaption}
\usepackage[colorlinks=true, linkcolor=blue, citecolor=blue, urlcolor=blue]{hyperref}
\hypersetup{
    colorlinks=true,
    linkcolor=blue,
    filecolor=magenta,      
    urlcolor=blue,
    pdftitle={Manuscript},
    pdfpagemode=FullScreen,
}

\usepackage[capitalize]{cleveref}
\crefformat{equation}{Eq.~(#2#1#3)} 

\usepackage{caption} 
\usepackage{ragged2e}
\DeclareCaptionJustification{myjustified}{\justifying}
\begin{document}

\newcommand{\NMSU}{Department of Physics, New Mexico State University, Las Cruces, NM 88001, USA}

\newcommand{\ang}{\text{\normalfont\AA}}
\newcommand{\dpar}{$d_{||}$ }
\newcommand{\dxzyz}{$(d_{\text{xz}}$, $d_{\text{yz}})$ }
\renewcommand{\thefootnote}{\fnsymbol{footnote}}
\newcommand{\red}{\textcolor{red}}
\newcommand{\blue}{\textcolor{blue}}

\title{Entropy-induced confinement in two-dimensional magnetic monopole gases}

\author{Prakash Timsina}
\affiliation{\NMSU}
\author{Boris Kiefer}
\affiliation{\NMSU}
\author{Ludi Miao*}
\affiliation{\NMSU}

\date{\today}


\begin{abstract}
\begin{center}
$^{*}$ Email: lmiao@nmsu.edu
\end{center}
Magnetic monopole quasiparticles in spin ice materials hold the potential for exploring new frontiers of physics that extend beyond Maxwell's equations. We have previously presented a two-dimensional magnetic monopole gas (2DMG) with a net charge, confined at the interface between spin-ice ($R_2$Ti$_2$O$_7$, $R$ = Dy, Ho) and antiferromagnetic iridate ($R_2$Ir$_2$O$_7$, $R$ = Dy, Ho), which is proposed to be driven by entropy. However, this mechanism needs to be verified and studied systematically. In this work, we quantitatively demonstrate that entropy is a key factor in the 2D confinement of the monopole gas. Starting from the nearest-neighbor interaction, we reveal that the competition between the entropy of spin-ice, which favors the 2D confinement, and the entropy of the monopoles' random walks, which favors the deconfinement, dictates the distribution of the monopoles within a few layers close to the interface. Our entropy-based monopole model accurately reproduces the monopole distribution obtained from the energy-based spin model, affirming that 2D confinement is entropy-driven. We further employ both models to show that the monopole distribution can be manipulated by an external magnetic field and temperature, holding promise for next-generation devices based on magnetic monopoles. In the presence of dipolar interactions, entropy continues to play a crucial role in controlling 2DMG behavior at finite temperatures. Our findings reveal the entropic mechanisms in 2DMG, enabling the manipulation of emergent quasiparticles at material interfaces.
\end{abstract}

\maketitle
\section*{Introduction}
Magnetic monopoles quasiparticles in spin-ice ($R_2$Ti$_2$O$_7$, $R$ = Dy, Ho) materials have attracted tremendous attentions \cite{Bramwell,Fennell,Morris,Bovo,Grams,Pan,Pan2,Dusad} due to their behavior beyond the Maxwell's equation and potential application in magnetic technologies. However, in bulk spin-ice, monopoles and  antimonopoles always appear in pairs, leading to a zero net magnetic charge of the system \cite{2DMG}. The inactive magnetic charge degree of freedom of the monopole systems lead to insulating properties for monopoles of the spin-ice system, which limits the exotic properties, which are well-known in the electron counterparts, such as superconductivity \cite{SRO-SC,RuO2-SC,FeSe-SC}, metal-to-insulator transition \cite{VO2_Nature,VO2_Science, NdNiO3, Ca2RuO4}, and topological insulating state \cite{QSHE, TI}. Recently, we proposed a magnetically charged two-dimensional magnetic monopole gas (2DMG) \cite{2DMG} at the interface between spin-ice ($R_2$Ti$_2$O$_7$, $R$ = Ho, Dy) \cite{Snyder, Ramirez} and antiferromagnetic pyrochlore iridates ($R_2$Ir$_2$O$_7$) \cite{Lefrançois}. The 2DMG is confined to a few unit cells at the interface and exhibits a monopole-metallic property, that resembles a 2D electron gas. In our previous study \cite{2DMG}, we proposed that entropy may cause the 2D confinement of the monopole gas. However, this assumption has not been verified and systematically studied.\\

Based on the previous hypothesis \cite{2DMG}, here we address this issue by creating an entropy-based model, which we will call \enquote{monopole model} for the rest of the text. In this model, we consider the monopole and spin ice entropy contributions, and systematically investigate their competition in confining the monopole gas at the interface. We then compare these results to the energy-based model on spins \cite{2DMG} (which we will call \enquote{spin model} for the rest of the text), and find agreement between them. This verifies that entropy is the key factor to control 2D monopole confinement and determining the thickness of the monopole layer in an otherwise energetically degenerate system. Additionally, we studied the effects of external magnetic fields and temperatures on the density and distribution of the monopole gas in monopole model, and found the control becomes increasingly sensitive as temperature decreases. Moreover, when dipolar interactions are considered, the influence of entropy still remains important in governing the behavior of 2DMG at finite temperatures.  These results establish entropy as a fundamental aspect of magnetic monopole physics, offering new opportunities to control and manipulate 2DMGs. Our study serves as a rare example of an entropy-based model describing system behavior, distinguishing it from most condensed matter physical problems where energy dominates the physics.

\section*{Results and Discussion}

\begin{figure*}
\begin{center}
\includegraphics[width=7.0in]{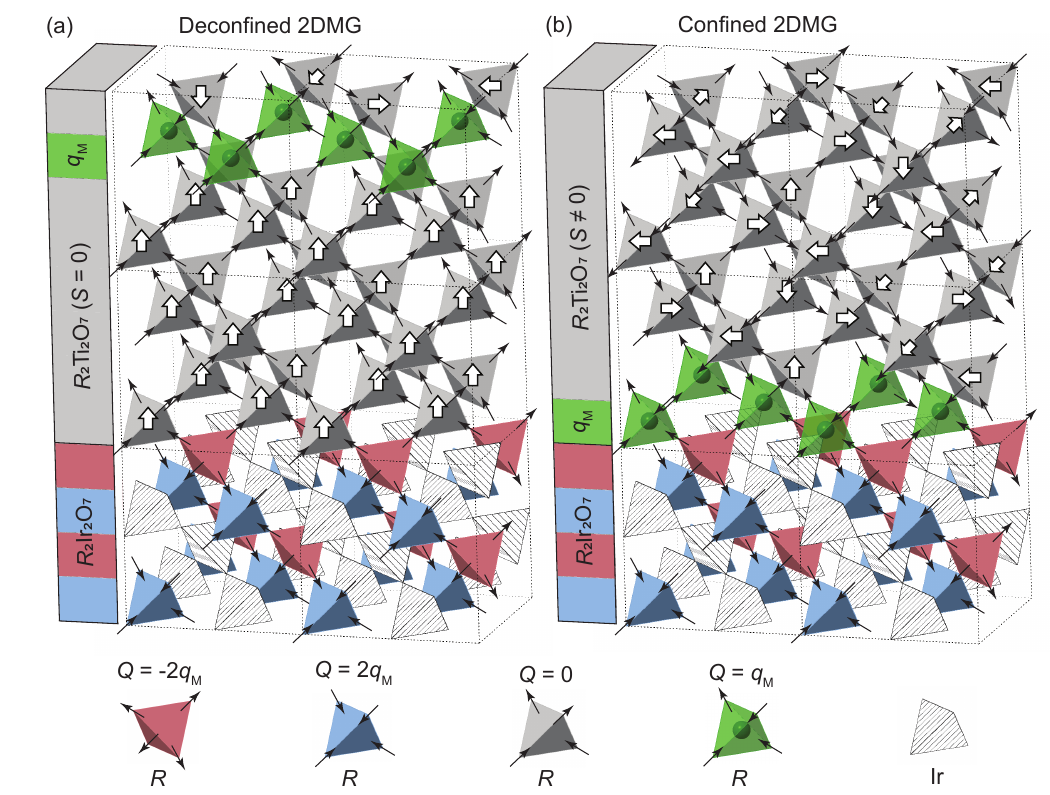}
\phantomsubcaption
\label{fig:one_a}
\phantomsubcaption
\label{fig:one_b}
\end{center}
\caption{Spin configurations of various monopole distributions in a $R_2$Ti$_2$O$_7$/$R_2$Ir$_2$O$_7$ heterostructure. Small black arrows represent the local moment of a rare earth ion, big white arrows represent the polarization direction of a tetrahedron. Tetrahedral sites are color-coded to show different magnetic charges: $Q$ = 0 for grey, $Q$ = $q_\text{M}$ for green, $Q$ = 2$q_\text{M}$ for blue, and $Q$ = -2$q_\text{M}$ for red. Spin and mononopole structures of two extreme cases are shown here: (a) all the monpoles live in the same layer, six layers off the interface, and (b) all the monopoles live in the same layer at the interface.}
\label{fig:one}
\end{figure*}

Spin-ice exhibits a \enquote{2-in-2-out} (2I2O) ice rule of spin configuration below 1 K \cite{Isakov, Snyder, Ramirez}. At finite temperatures, violation of the ice rules leads to \enquote{3-in-1-out/1-in-3-out} (3I1O/1I3O) defects, also known as magnetic monopole quasi-particles \cite{Castelnovo2}. In a nearest-neighbor interaction model, monopole hopping is an energetically degenerate process. This energetic equivalence grants the monopole the freedom to move freely within spin-ice. In an  $R_2$Ti$_2$O$_7$/$R_2$Ir$_2$O$_7$ heterostructure, a monopole gas of the same sign of charge exists in the spin-ice $R_2$Ti$_2$O$_7$ layer \cite{2DMG}. Intuitively, one might expect a uniform distribution of monopoles across the spin-ice layer. However, our Monte Carlo simulation reveals that monopoles are concentrated near the interface of the heterostructure \cite{2DMG}, whose origin needs to be understood.\\

To address this puzzle, we consider two extreme cases of spin configurations, one where the monopoles occupy a single tetrahedral layer, six layers away from the interface Fig. \ref{fig:one_a}; and another where monopoles occupy a single layer at the interface Fig. \ref{fig:one_b}. Though these configurations have identical total energies in a nearest-neighbor interaction model, their entropies differ. In Fig. \ref{fig:one_a}, only the fully polarized spin configuration is permissible between the interface and the monopole layer, resulting in zero entropy due to the interface's boundary conditions and the absence of monopoles in these layers. Conversely, in the scenario depicted in Fig. \ref{fig:one_b}, where monopoles are confined to the interface, the same region can adopt numerous 2I2O spin configurations, leading to the well-known zero-point entropy of the spin-ice \cite{Ramirez}. Comparing these cases, the one depicted in Fig. \ref{fig:one_b} has a higher entropy than Fig. \ref{fig:one_a}, which is equal to the spin-ice zero-point entropy for the region between the interface and the monopole layer in Fig. \ref{fig:one_a}. Consequently, the configuration with monopoles confined at the interface emerges as the more probable one.

\begin{figure*}
\begin{center}
\includegraphics[width=7.0in]{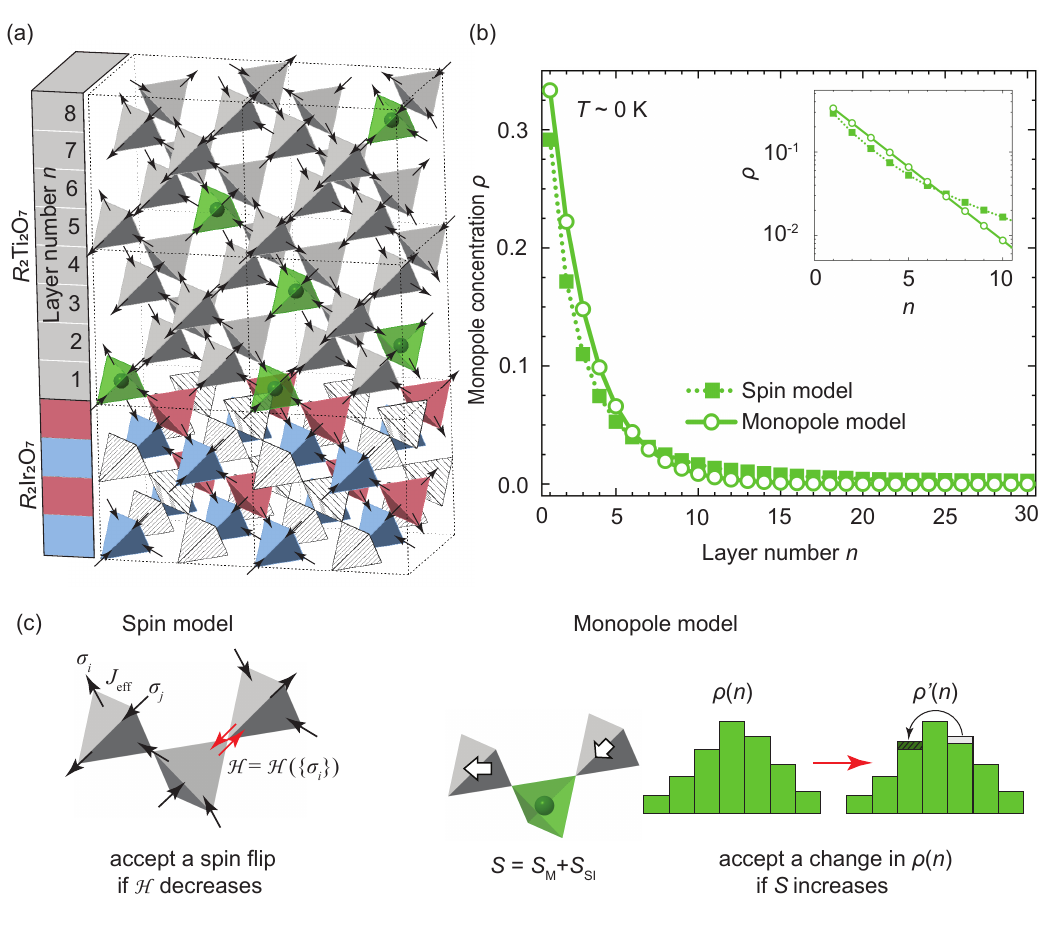}
\phantomsubcaption
\label{fig:two_a}
\phantomsubcaption
\label{fig:two_b}
\phantomsubcaption
\label{fig:two_c}
\end{center}
\caption{Monopole distribution at $T \sim 0\,\text{K}$. (a) Illustration of monopole distribution in a $R_2$Ti$_2$O$_7$/$R_2$Ir$_2$O$_7$ heterostructure. More monopoles are concentrated near to interface. (b) Monopole concentration $\rho$ as a function of layer number $n$ of the 2DMG at $T \sim 0\,\text{K}$ from the spin Monte Carlo simulation and from the entropy maximization. (c) Schematics of Monte Carlo simulation on spin model, and entropy maximization on the monopole model, both are performed at the $T \sim 0\,\text{K}$. $\mathscr{H}$ is the energy of the system as defined in \cref{Eq:twoadded}.}
\label{fig:two}
\end{figure*}

To understand the monopole distributions, it is essential to consider not only the entropy contribution from spin-ice but also the entropy associated with monopole random walks. While spin-ice entropy tends to push monopoles toward the interface as discussed in Fig. \ref{fig:one}, the monopole entropy drives monopoles into random walks, dispersing the monopoles throughout the spin-ice layers. This competition between the two entropy forces results in a distribution where monopoles are predominantly found near the interface, as shown in Fig. \ref{fig:two_a}.

To accurately analyze the result of this competition, we look for an optimized monopole distribution function $\rho (n)$, where $\rho$ is the monopole concentration ($\rho=1$ for a layer full of monopoles and $\rho =0$ for a layer with no monopoles) and $n$ is the tetrahedral layer number counting from the interface so that the total entropy $S[\rho(n)]$ maximizes among all possible $\rho(n)$. The total entropy $S[\rho(n)]$ at $T \sim 0\,\text{K}$ can be written as:
\begin{equation}
\label{Eq:one}
S[\rho(n)]=-\sum_{n} k_\text{B} \rho(n)\ln \rho(n)+\sum_{n}2S_0 [1-M(n)]
\end{equation}

where $S_0 = k_\text{B} 1/2 \ln\left(3/2\right)$ is the zero-point entropy of spin-ice per spin \cite{Ramirez}, $M(n) = 1 - \sum_{m<n} \rho(m)$ is the normalized polarization of each layer ($M=1$ for a fully polarized tetrahedral layer and $M=0$ for a non-polarized tetrahedral layer). The first term is the entropy contribution from the monopole random walks \cite{Gao}. The second term is the entropy contribution from the spin-ice \cite{Ramirez}. We calculated the monopole distribution function $\rho(n)$ by maximizing the total entropy, as shown in Fig. \ref{fig:two_c}. The result is shown in Fig. \ref{fig:two_b}. The monopole concentration $\rho$ is highest in the first tetrahedral layer (around 1/3), which monotonously decreases toward the interior of the spin-ice, to $\sim 10\%$ of the first layer, over the first eight tetrahedral layers. \\

We performed Monte Carlo simulations on spin model \cite{2DMG} as shown in Fig. \ref{fig:two_c}, an independent method from the monopole model, to investigate the monopole distribution near the $R_2$Ti$_2$O$_7$/$R_2$Ir$_2$O$_7$ interface at $T \sim 0\,\text{K}$. 

The Hamiltonian is given by \cite{2DMG}:

\begin{equation}
\label{Eq:twoadded}
\mathscr{H} = J_{\text{eff}} \sum_{\langle i,j \rangle} \sigma_i \sigma_j + \frac{1}{6} H_{\text{loc}} \sum_{\langle i,\alpha \rangle} \sigma_i \sigma_{\alpha}
\end{equation}

where $\sigma_i$, and  $\sigma_j$ (both taking values $\pm$1) represent the Ising pseudo-spins of $R^{3+}$ and $\sigma_{\alpha}$ (taking values $\pm$1) indicates the Ising pseudo-spins of Ir$^{4+}$, oriented towards or away from a tetrahedron. The notation $\langle i, j \rangle$ denotes summation over nearest-neighbor sites. The parameter $J_{\text{eff}}$ represents the effective nearest-neighbor interaction between $R^{3+}$ moments. And, $H_{\text{loc}}$ is a local background static field generated by Ir$^{4+}$ moments over the $R^{3+}$ moments.
In the Monte Carlo simulation, each time a random spin associated with $R^{3+}$ is visited, and the spin flip is attempted. The flip is always allowed when the energy is favorable, and partially allowed when the energy is unfavorable by the probability $e^{-{\Delta E}/{k_{\text{B}} T}}$, where $\Delta E$ is the energy cost due to the spin flip. The moments associated with Ir$^{4+}$ ions are considered only for the local static background field which are static during the Monte Carlo process.\\

At absolute zero temperature, the trilayer system of $R_2$Ir$_2$O$_7$/$R_2$Ti$_2$O$_7$/$R_2$Ir$_2$O$_7$ generates a non-zero exchange field from both the $R_2$Ir$_2$O$_7$/$R_2$Ti$_2$O$_7$ and $R_2$Ti$_2$O$_7$/$R_2$Ir$_2$O$_7$ interfaces. This field induces a flux that either flows towards or away from the spin ice, leading to the formation of 2DMG \cite{2DMG}. In contrast, in the single interface case of $R_2$Ti$_2$O$_7$/$R_2$Ir$_2$O$_7$, the absolute 0 K state does not exist and the 2DMG can only form at arbitrarily low temperatures, which is defined as $T_\epsilon$. The state corresponding to $T_\epsilon$ is referred as ``quasi-ground state''. In the texts and figures, we used $T \sim 0\,\text{K}$ for $T = T_\epsilon$. Details on the stableness of 2DMG and quasi-ground states at a single interface of $R_2$Ti$_2$O$_7$/$R_2$Ir$_2$O$_7$ are discussed in Supplementary Note 1 \cite{SupplementaryInformation}.\\

In the spin model, we obtained monopole distribution function $\rho(n)$ by taking an average of 4,000 spin configuration snapshots, as shown in Fig. \ref{fig:two_b}. We have varied the system sizes to make sure that the result is independent to the system size (see Supplementary Note 2 \cite{SupplementaryInformation}). Remarkably, the monopole distribution function obtained by entropy maximization in monopole model and spin model agree very well (Fig. \ref{fig:two_b}). This indicates that entropy plays a pivotal role in governing the distribution of monopoles within the 2DMG system.

Results from both methods exhibit a very roughly exponential behavior in monopole concentration $\rho$ with respect to the number of layers $n$, as illustrated in Fig. \ref{fig:two_b} (inset). The exponential behavior of $\rho$ can be understood as a mathematical consequence of entropy maximization (see Supplementary Note 3 \cite{SupplementaryInformation} for details). The slight discrepancy between the results of the monopole and spin models in Fig. \ref{fig:two_b} (inset) is attributed to the approximation made in the entropy expression of the monopole model (see Supplementary Note 4 \cite{SupplementaryInformation} (see also reference \cite{Cathelin2} therein) for detailed explanation).
\begin{figure*}
\begin{center}
\includegraphics[width=7.0in]{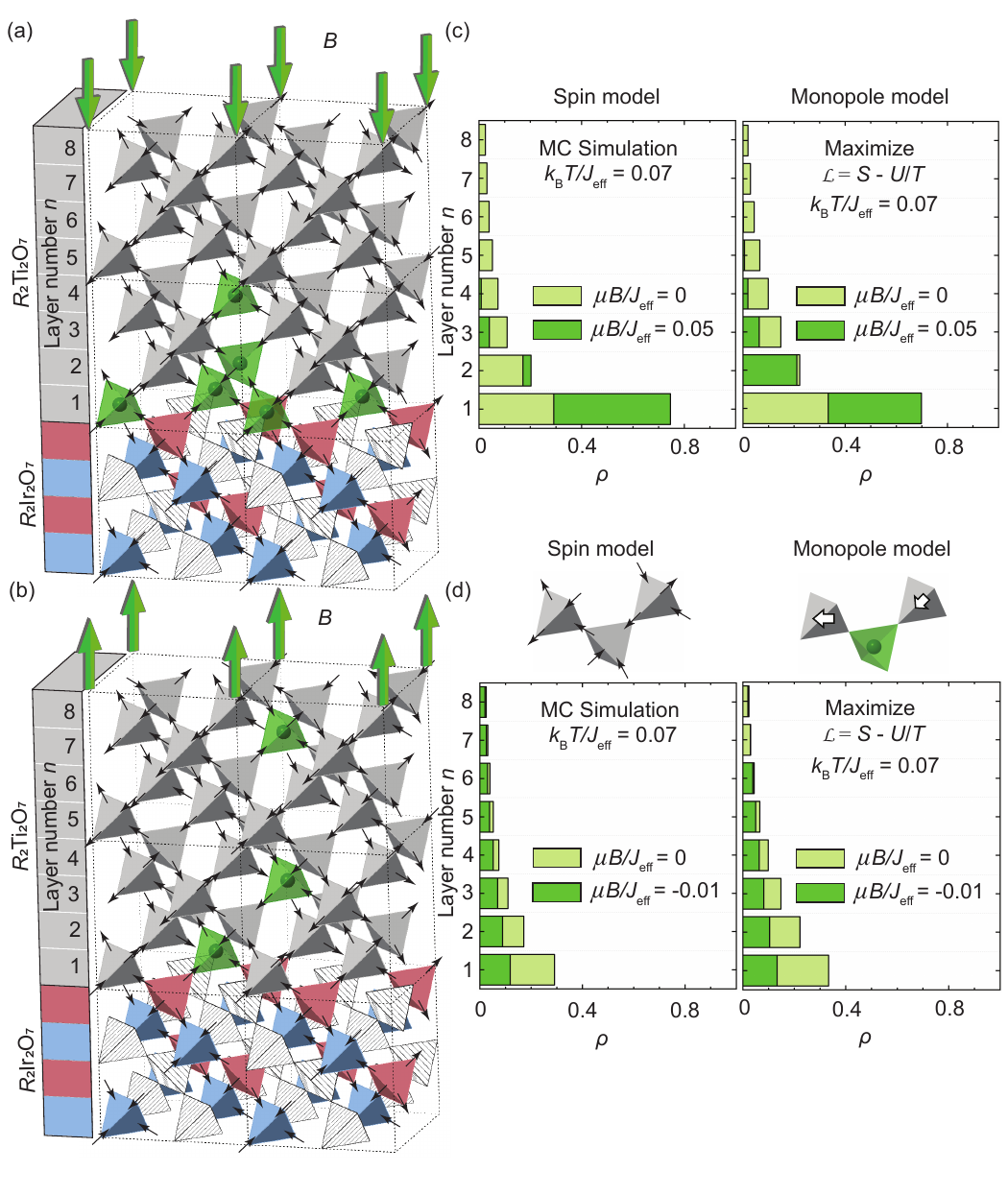}
\phantomsubcaption
\label{fig:three_a}
\phantomsubcaption
\label{fig:three_b}
\phantomsubcaption
\label{fig:three_c}
\phantomsubcaption
\label{fig:three_d}
\end{center}
\caption{Monopole distribution at finite temperature with magnetic field. (a) and (b) Illustration of monopole distribution in a $R_2$Ti$_2$O$_7$/$R_2$Ir$_2$O$_7$ heterostructure with magnetic field point toward the iridate side and titanate side, respectively. (c) and (d) Monopole distribution function $\rho(n)$ calculated by spin Monte Carlo simulation and modified entropy maximization, at $T$ = 0.1 K (\( k_{\text{B}}T / J_{\text{eff}} = 0.07 \)), with a field of 20 mT ($\mu B / J_{\text{eff}} = 0.05$) (point toward iridate side) and -4 mT ($\mu B / J_{\text{eff}} = -0.01$) (point toward titanate side), respectively. The calculation results without field at quasi-ground state for each model are overlayed with light green color for comparison.}
\label{fig:three}
\end{figure*}

Having understood the monopole distribution at the quasi-ground state without magnetic field, we now turn to explore on how these external perturbations can provide additional knobs to control the system. 
We applied a finite temperature $T$ and an external magnetic field $B$ perpendicular to the interface and investigated their impact on the monopole distributions. At finite $T$, a system should possess a minimal free energy $F(B,T) = U(B)-TS$. 
where $U(B)$ is the total energy of the system in a magnetic field $B$. Therefore, in searching for a monopole distribution $\rho(n)$ with a minimum $F(B,T)$, we maximize $L(B,T)=S-U(B)/T$, where $L(B,T)$ is the modified entropy under finite temperature and magnetic field, defined by:

\begin{equation}
\label{Eq:two}
L[\rho(n);B,T] = S[\rho(n)] -\sum_{n}  k_\text{B}  q_\text{M} \rho(n) n a B /T    
\end{equation}

where $S$ is the total entropy as described in \cref{Eq:one}, $q_\text{M} =(8/\sqrt3)\mu/a$ is the magnetic charge of a monopole quasiparticle \cite {Castelnovo2, 2DMG}, $\mu\sim10 \, \mu_\text{B}$ is the $R^\text{3+}$ moment and $a$ is the pyrochlore lattice constant. The second term is the magnetic potential energy of the monopoles in a magnetic field, then divided by temperature.

We set $T$ = 0.1 K ($k_{\text{B}}T / J_{\text{eff}} = 0.07$), where $J_{\text{eff}}$ is the nearest exchange interaction between two rare-earth local moments and $B$ = 20 mT ($\mu B / J_{\text{eff}} = 0.05$) (the positive magnetic field direction is defined as along out-of-plane direction and pointing toward the iridate side) and obtained a monopole distribution $\rho(n)$ with maximized modified entropy $L$, as shown in Fig. \ref{fig:three_c}, labeled as monopole model. The monopole model calculation result without field at quasi-ground state is shown in the same panel for comparison. Under a positive magnetic field, the monopoles are more confined at the interface: with the same total sheet density $\Sigma_n \rho(n)=1$. To investigate the effect of the negative field (which is defined as along out-of-plane direction and pointing toward the titanate side), we set $T$ = 0.1 K ($k_{\text{B}}T / J_{\text{eff}} = 0.07$) and $B$ = -4 mT ($\mu B / J_{\text{eff}} = -0.01$), and find a $\rho(n)$ with maximized $L$, as shown in Fig. \ref{fig:three_d}. The negative field energetically drives some monopoles away from the interface, so that $\Sigma_n \rho(n)<1$. To evaluate these results, we performed Monte Carlo simulations on spin model \cite{2DMG} at the same temperature and magnetic field, as shown in Figs. \ref{fig:three_c} and \ref{fig:three_d}, labeled as spin model. Again the monopole distributions obtained by both approaches agree very well, indicating that under magnetic fields at finite temperatures, total energy and total entropy both play important roles in determining the monopole behavior, which provides the opportunities of manipulating monopoles using a conventional energetic approach as well as exotic entropic approach.

\begin{figure*}
\begin{center}
\includegraphics[width=7.0in]{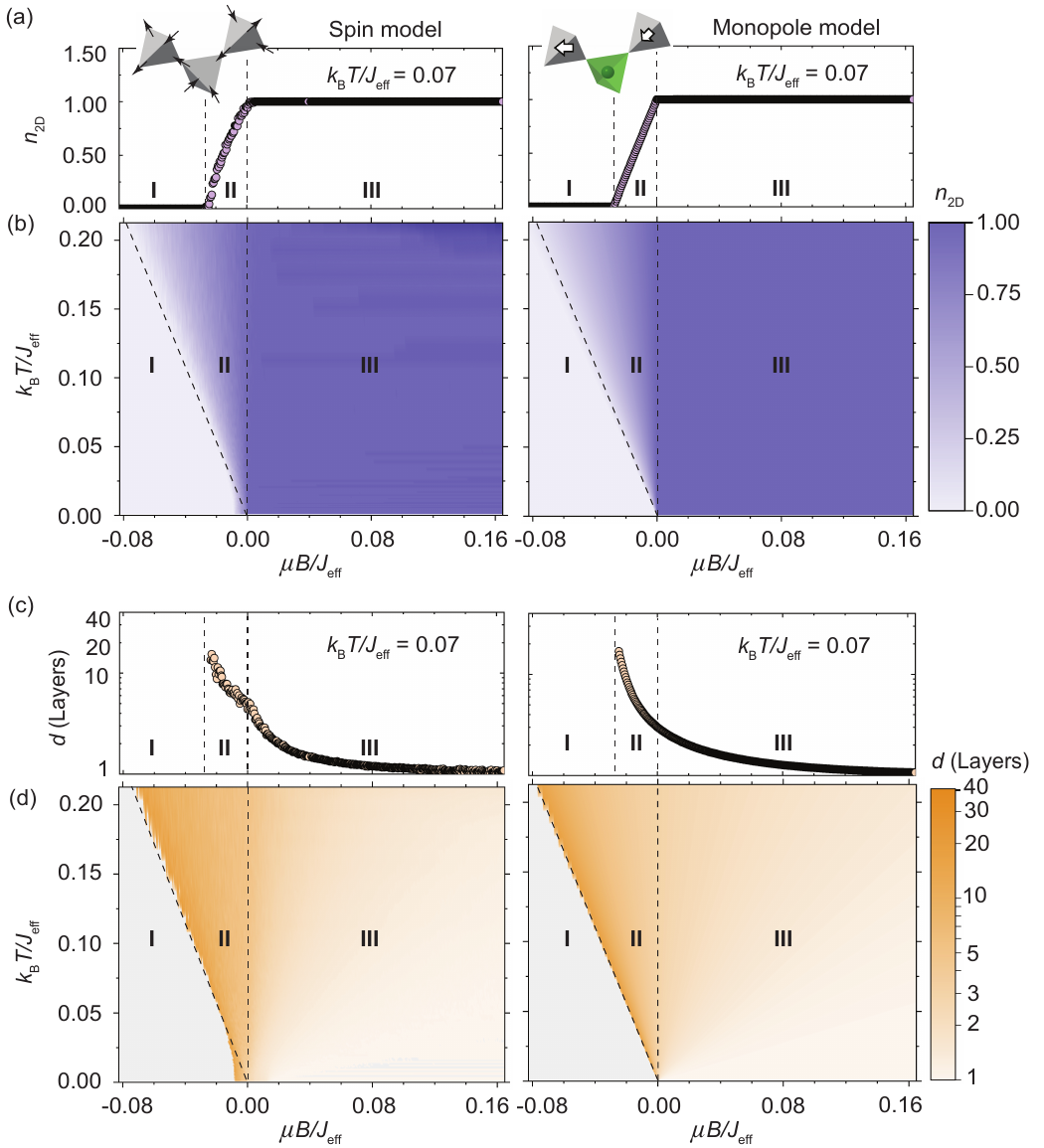}
\phantomsubcaption
\label{fig:four_a}
\phantomsubcaption
\label{fig:four_b}
\phantomsubcaption
\label{fig:four_c}
\phantomsubcaption
\label{fig:four_d}
\end{center}
\caption{Temperature-magnetic field diagram of the 2DMG. (a) Interface monopole sheet density $n_\text{2D}$ defined as $n_\text{2D} = \sum_{n} \rho(n)$, as a function of out-of-plane relative magnetic field $\mu B / J_{\text{eff}}$ calculated by Monte Carlo simulation on spin model and entropy maximization method at $T$ = 0.1 K (\( k_{\text{B}}T / J_{\text{eff}} = 0.07 \)). (b) $n_\text{2D}$ as a map of $T$ and $B$ calculated by both methods. (c) Interface monopole depth defined as $d = \sum_{n} n\rho(n)/\sum_{n} \rho(n)$ as a function of $\mu B / J_{\text{eff}}$ calculated by both method at $T$ = 0.1 K (\( k_{\text{B}}T / J_{\text{eff}} = 0.07 \)). (d) $d$ as a map of $T$ and $B$ calculated by both methods. Three regions are labeled as I, II, and III. Dashed lines represent boundaries between different regions.}
\label{fig:four}
\end{figure*}

The distribution of monopoles can be quantitatively described by two quantities: interface monopole sheet density defined as $n_\text{2D} = \sum_{n} \rho(n)$ and the interface monopole depth defined as $d = \sum_{n} n\rho(n)/\sum_{n} \rho(n)$. Here, we reconcile the summation of $n_\text{2D}$ and $d$ over half of the system size to exclude the effect of unwanted surface on the other side of the interface (see Supplementary Note 5 \cite{SupplementaryInformation}). Figures \ref{fig:four_a} and \ref{fig:four_c} show $n_\text{2D}$ and $d$ as a function of magnetic field $B$ calculated at $T$ = 0.1 K ($k_{\text{B}}T / J_{\text{eff}} = 0.07$) using both methods, respectively. We have tentatively assigned three regions with distinct behaviors: region I, $\mu B / J_{\text{eff}} < -0.027$ , where all the monopoles are pumped away by the magnetic field with $n_\text{2D}=0$ and $d$ ill-defined; region II, $-0.027 < \mu B / J_{\text{eff}} < 0$, where a less confined ($d>3$) and lower density ($n_\text{2D}<1$) monopole gas appear near the interface; and region III, $\mu B / J_{\text{eff}} > 0$ , where a more confined ($d<3$) and constant density ($n_\text{2D}=1$) monopole gas appear at the interface. Further increasing the magnetic field on the positive side will pull all the monopoles in the first layer. Decrease in $n_\text{2D}$ and deconfinement of monopoles observed during the transition from region II to region I of Fig. \ref{fig:four} is indicative of Kasteleyn transition \cite{Jaubert, Pili} in our
2DMG system (see Supplementary Note 6 \cite{SupplementaryInformation} for details).

Having demonstrated the roles of temperature and magnetic field, we expand the above two-method calculations and analysis to a T-B diagram up to $T$ = 0.3 K ($k_{\text{B}}T / J_{\text{eff}} = 0.21$) (where antimonopoles start to appear due to thermal excitation), as shown in Figs. \ref{fig:four_b} and \ref{fig:four_d}. In the T-B diagram, all these regions exist at all temperatures. The boundaries of different regions converge at a critical point (CP) $T = 0$ K and $B = 0$ T at low temperatures, indicating the system is very sensitive to a magnetic field close to the CP. The low-temperature field sensitivity can be understood as follows: a factor of $B/T$ is the only place $B$ and $T$ appear in the \cref{Eq:two}, where $B$ and $T$ scales linearly in $L(B,T)$. Since the transition is determined by the $B/T$ value, the field requirement for the transition should be smaller at lower temperatures. The low-temperature field sensitivity and the CP of the 2DMG T-B diagram is a consequence of an energetically degenerate process of monopole hopping in spin-ice.

The above results consider only nearest-neighbor interactions, which provides valuable insights into the impact of entropy on monopole behavior in other models. For example, in the model where long-range dipolar interactions are considered \cite{Castelnovo2, Hertog, Melko}, entropy still plays an important role in deciding the monopole behavior: the total entropy is still contributed by spin-ice entropy and monopole entropy of random walks, whereas a monopole-monopole interaction should be added to the total energy. With long-range interaction, the modified entropy expression becomes,
\begin{equation}
\label{eq:three}
L[\rho(n);T] = S[\rho(n)] -U/T    
\end{equation}

where $S$ is the total entropy from \cref{Eq:one}, $U =  \frac{\mu_0}{4\pi} \frac{Q_\alpha Q_\beta}{r_{\alpha \beta}}$ is the magnetic Coulomb energy between two monopoles $Q_\alpha$ and $Q_\beta$ \cite{Castelnovo2}, $r_{\alpha \beta}$ is the distance between two monopoles, $Q_\alpha$ is the total magnetic charge at site $\alpha$. The expression \cref{eq:three} contains long-range dipolar interaction energy, $E_d = Dl^3\sum_{(ij)}\left[\frac{\hat{e}_i\cdot\hat{e}_j}{|\mathbf{r}_{ij}|^3}-\frac{3(\hat{e}_i\cdot\mathbf{r}_{ij})(\hat{e}_j\cdot\mathbf{r}_{ij})}{|\mathbf{r}_{ij}|^5}\right]\sigma_i \sigma_j$ \cite{Castelnovo2}, \(D = \mu_0\mu^2/(4\pi l^3)\) is the coupling constant of the dipolar interaction, $l$ is the pyrochlore nearest-neighbor distance, \(\sigma_i\) and \(\sigma_j\) are the Ising spin variables, and \(\mathbf{r}_{ij}\) is the distance between spins \(i\) and \(j\). We used a stabilized spin configuration resulting from monopole repulsion due to dipolar interactions and calculated the dipolar interaction energy $\Delta U$ as $k_{\text{B}}$0.275 K. We then estimated the effective temperature of the entropic effect, where in the free energy $\Delta F = \Delta U - T \Delta S$, the system entropy change $\Delta S$ times $T$ should be at least 10$\%$ of the dipolar interaction energy scale to be non-negligible. Using $\Delta S = \frac{1}{2} k_{\text{B}} \ln\left(\frac{3}{2}\right)$ and $\Delta U$ = $k_{\text{B}}$0.275 K for energy scale estimation, we find $T>$ 0.14 K. Consequently, the entorpy begins to correct the distribution of monopoles at 0.14 K and becomes significant when $T>$ 0.14 K, even in the presence of long-range dipolar interactions (see Supplementary Note 7 \cite{SupplementaryInformation} for details).

Our results provide insights into entropy-related characterization in the 2DMG system. For example, monopoles do not carry electrical charge but carry energy and entropy. Therefore, thermal transport measurement \cite{Grams, Pruschke, Suszalski,RuCl3-1,RuCl3-2} of the heterostructure would be a viable method to characterize the monopole behaviors. Compared to electron systems such as 2D electron gases \cite{Ohtomo} which is governed by energy, entropy plays an important role in 2DMG. 

\section*{Conclusion}
In this study, we systematically investigated the physical principles that govern the two-dimensional (2D) confinement of monopole gases at spin ice ($R_2$Ti$_2$O$_7$, $R$ = Dy, Ho) and antiferromagnetic iridate ($R_2$Ir$_2$O$_7$, $R$ = Dy, Ho) interface. Our results show that in the nearest-neighbor model, this confinement is primarily driven by entropic effects. We identified the significant contributions of entropy from both spin-ice degenerated states and the monopole gas itself through the entropy-based monopole model, which compares well with the energy-based spin model. These contributions lead to the localization of the two-dimensional monopole gas (2DMG) in the vicinity of the interface, a phenomenon that occurs despite the absence of energy cost/gain for monopole to hop between lattice sites. We further explore how the density and spatial distribution of the 2DMG can be manipulated through the application of external magnetic fields and temperatures. Notably, we find that 2DMG's response to these external stimuli becomes markedly more sensitive at lower temperatures, especially in the vicinity of $ T = 0$ K, $B = 0$ T, a critical point. In the dipolar interactions, entropy still plays an important role in 2DMG behavior at finite temperatures ($T > $ 0.14 K). The insights gained from this study are promising to guide the thermal transport measurement for detecting the 2DMG, as well as the development of next-generation devices based on magnetic monopoles.

\section*{METHODS}
\textbf{Entropy Variational Method} The monopole picture simulations are performed by modeling the entropy contributions from both the spin ice lattice and the monopole gas. The total entropy is given by  \cref{Eq:one}, which incorporates the inherent monopole entropy $- \sum_{n} k_\text{B} \rho \ln(\rho)$ \cite{Gao} that depends on the monopole density $\rho$ in each layer, as well as the spin ice entropy $\sum_{n} 2S_o [1 - M(n)]$, where $S_0 = k_\text{B} 1/2\ln\left(3/2\right)$ represents the zero-point entropy of spin ice per spin \cite{Ramirez}, that relies on the polarization factor $M(n)$ calculated from monopole distributions, with $M=1$ corresponding to a fully polarized tetrahedral layer and $M=0$ to a non-polarized one. To find the optimum monopole density distribution $\rho$ that maximizes the total entropy, we search for $\rho$ values that give the maximum entropy $S$. We also model the system under an external magnetic field $B$ and a finite temperature $T$ by using the modified entropy expression in  \cref{Eq:two}. This additional entropy term, $- \sum_{n} k_\text{B} q_\text{M} \rho(n)na B/T$, captures the effects of the external perturbations. By maximizing this modified entropy $L$, we determine the optimal monopole distribution $\rho$ under different $B$ and $T$. The monopole densities $\rho$ are initialized randomly, and the variational method with constraint $\sum_{n} \rho(n) = 1$, optimizes $\rho$ values to maximize entropy. Convergence is determined when changes in entropy fall below a convergence threshold of $10^{-9}$. We used 75 and 107 crystal layers for the quasi-ground state and in the external perturbation case, respectively.  Monopole model simulations are implemented in C++ using numerical optimization of the entropy expressions. 

\textbf{Monte Carlo Simulation} The spin model based on Monte Carlo simulations is implemented to explore all possible spin configurations and determine the most probable spin state. The simulations account for nearest-neighbor exchange interactions between the spins. A single spin-flip Metropolis algorithm \cite{Lefrançois} with periodic boundary conditions is utilized, where spins are randomly flipped to find the favorable state. The simulations were performed on $4 \times 4 \times 30$
lattice sites with periodic boundary conditions for $R_2$Ti$_2$O$_7$/$R_2$Ir$_2$O$_7$ heterostructure in external field case and $4 \times 4 \times 20$ lattice sites without field. We have around 1000 thermalization steps for approximately each $10^{5}$ single spin-flip step. The output provides spin configurations, from which monopole distributions are deduced which maps to magnetic charges. Further implementation details of the Monte Carlo algorithm can be found in previous work \cite{2DMG}. This Monte Carlo simulation is implemented in C++.

\section*{Data and  code availability}
All relevant data and code are available from the authors upon reasonable request.


\section*{Competing Interests}
The authors declare no competing interests

\end{document}